\documentclass[11pt]{article}
\usepackage{amsfonts}
\usepackage{amsmath}
\usepackage{epsfig}

\begin{document}

\begin{center}
{\Large Persistent random walk of cells involving anomalous effects and
random death }

{\large Sergei Fedotov}$^{1}${\large , Abby Tan}$^{2}${\large \ and Andrey
Zubarev}$^{3}$

\bigskip

$^{1}$School of Mathematics, The University of Manchester, UK;

$^{2}$Department of Mathematics, Universiti Brunei Darussalam, Brunei;

$^{3}$Department of Mathematical Physics, Ural Federal University,
Yekaterinburg, Russia.

\bigskip

\bigskip

\textbf{Abstract}
\end{center}

The purpose of this paper is to implement a random death process into a
persistent random walk model which produces subballistic superdiffusion (L%
\'{e}vy walk). We develop a Markovian model of cell motility with the extra
residence variable $\tau .$ The model involves a switching mechanism for
cell velocity with dependence of switching rates on $\tau $. This dependence
generates intermediate subballistic superdiffusion. We derive master
equations for the cell densities with the generalized switching terms
involving the tempered fractional material derivatives. We show that the
random death of cells has an important implication for the transport process
through tempering of superdiffusive process. In the long-time limit we write
stationary master equations in terms of exponentially truncated fractional
derivatives in which the rate of death plays the role of tempering of a L%
\'{e}vy jump distribution. We find the upper and \ lower bounds for the
stationary profiles corresponding to the ballistic transport and diffusion
with the death rate dependent diffusion coefficient. Monte Carlo simulations
confirm these bounds.

\section{ Introduction}

Cell motility is an important factor in embryonic morphogenesis, wound
healing, cancer proliferation, and many other physiological and pathological
processes \cite{RRR}. The microscopic theory of the cell migration is based
on various random walk models \cite{EO}. Most theoretical studies of cell
motility deal with Markovian random walks \cite{Alt,sad2,O}. However, the
experimental analysis of the trajectories of cells shows that they might
exhibit non-Markovian superdiffusive dynamics \cite{J,PK,Jap}. It has been
found recently that cancer cells motility is superdiffusive \cite%
{Mierke,Mierke2}.

Several techniques are available to obtain a superdiffusion including the
continuous time random walk (CTRW) \cite{K,MK,Z}, generalization of the
Markovian persistent random walk \cite{Mas,Fer,Sok}, stochastic differential
equations \cite{West}, a fractional Klein--Kramers equation \cite{PK,Barkai}%
, non-Markovian switching model \cite{FMT}. The CTRW model \cite{K,MK,Z} for
superdiffusion involves the joint probability density function (PDF) $\Phi
(\tau ,r)$ for a waiting time $\tau $ and a displacement (jump) $r$. One has
to assume that the waiting time and displacement are correlated. For
example, $\Phi (\tau ,r)$ $=\delta (\tau -|r|/v)w(r)$ with $w(r)$ $\sim
|r|^{-\mu }$ ($2<\mu <3)$ as $r\rightarrow \infty $ corresponds to the L\'{e}%
vy walk for which the particle moves with a constant speed $v$, and the
waiting time $\tau $ depends on the displacement. The mean-square
displacement for the L\'{e}vy walk is $\mathbb{E}X^{2}(t)\sim t^{4-\mu }$
(superdiffusion). Another way to obtain a superdiffusive bevaviour is a
two-state model with power law sojourn time densities as the generalization
of correlated random walk involving two velocities \cite{Mas,Fer,Sok}. One
can \ also start with the stochastic differential equation for the position
of particle $X(t):$ $\dot{X}(t)=v(t),$ where the velocity $v(t)$ is a
dichotomous stationary random process with zero mean which takes two values $%
V$ and $-V$ \cite{West}. One can obtain a superdiffusive increase of the
mean squared displacement in time by using a fractional Klein--Kramers
equation for the probability density function for the position and velocity
of cells \cite{PK,Barkai}. This equation generates power law velocity
autocorrelation, $C_{v}(t)\sim E_{\mu }\left[ -\left( \frac{t}{\tau _{0}}%
\right) ^{\mu }\right] ,$ involving the Mittag-Leffler function $E_{\mu }$
which explains the superdiffusive behaviour. In \cite{FMT} the authors
proposed a Markov model with an ergodic two-component switching mechanism
that dynamically generates anomalous superdiffusion.

In this paper we address the problem of the mesoscopic description of
transport of cells performing superdiffusion with the random death process.
One of the main challenges is how to implement the death process into a
non-Markovian transport processes governed by a persistent random walk with
power law velocity autocovariance. We do not impose the power-law velocity
correlations at the very beginning. Rather, this correlation function is
dynamically generated by internal switching involving the age dependent
switching rate. There exist several approaches and techniques to deal with
the problem of persistent random walk with reactions \cite%
{MFH,Had,H,Hi,Bo,Mend}. However these works are concerned only with a
Markovian switching between two states. Our main objective here is to
incorporate the death process into non-Markovian superdiffusive transport
equations which is still an open problem. We show that the random death of
cells has an important implication for the transport process through
tempering of superdiffusive process.

\section{ Persistent random walk model\ involving superdiffusion}

The basic setting of our model is as follows. The cell moves on the right
and left with the constant velocity $v$ and turns with the rate $\gamma
(\tau )$. The essential feature of our model is that the switching rate $%
\gamma (\tau )$ depends upon the time which the cell has spent moving in one
direction \cite{Alt}. We suggest that the switching rate $\gamma (\tau )$ is
a decreasing function of residence time $\tau $ (negative aging). This rate
describes the anomalous persistence of cell motility: the longer cell moves
in one direction, the smaller the switching probability to another direction
becomes. Keeping in mind a superdiffusive movement of the cancer cells \cite%
{Mierke,Mierke2}, we consider the inhibition of cell proliferation by
anticancer therapeutic agents \cite{Iomin}. To describe this inhibition we
consider the random death process assuming that during a small time interval
$\left( t,t+\Delta t\right) $ each cell has a chance $\theta \Delta
t+o(\Delta t)$ of dying, where $\theta $ is the constant death rate. In what
follows we show that the governing equations for the cells densities involve
a non-trivial combination of transport and death kinetic terms because of
memory effects \cite{MFH,Abad,Zub,Ang,Steve}.

Let us define the mean density of cells, $n_{+}(x,t,\tau ),$ at point $x$
and time $t$ that move in the right direction with constant velocity $v$
during time $\tau $ since the last switching. The mean density $%
n_{-}(x,t,\tau )$ corresponds to the cell movement on the left. The balance
equations for both densities $n_{+}(x,t,\tau )$ and $n_{-}(x,t,\tau )$ can
be written as
\begin{equation}
\frac{\partial n_{+}}{\partial t}+v\frac{\partial n_{+}}{\partial x}+\frac{%
\partial n_{+}}{\partial \tau }=-\gamma (\tau )n_{+}-\theta n_{+},
\label{str}
\end{equation}%
\begin{equation}
\frac{\partial n_{-}}{\partial t}-v\frac{\partial n_{-}}{\partial x}+\frac{%
\partial n_{-}}{\partial \tau }=-\gamma (\tau )n_{-}-\theta n_{-},
\label{str2}
\end{equation}%
where $\gamma (\tau )$ is the switching rate and $\theta $ is the constant
death rate. We assume that at the initial time $t=0$ all cells just start to
move such that%
\begin{equation}
n_{\pm }(x,0,\tau )=\rho _{\pm }^{0}(x)\delta (\tau ),  \label{initial}
\end{equation}%
where $\rho _{+}^{0}(x)$ and $\rho _{-}^{0}(x)$ are the initial densities.

Our aim is to derive the master equations for the mean density of cells
moving right, $\rho _{+}(x,t),$ and the mean density of cells moving left, $%
\rho _{-}(x,t)$ defined as%
\begin{equation}
\rho _{\pm }(x,t)=\int_{0}^{t^{+}}n_{\pm }(x,t,\tau )d\tau ,  \label{den}
\end{equation}%
where the upper limit of $t^{+}$ is shorthand notation for $%
\lim_{\varepsilon \rightarrow 0}\int_{0}^{t+\varepsilon }.$ This limit
emphasizes that singularity located at $\tau =t$ is entirely captured by the
integration with respect to the residence variable $\tau $. Boundary
conditions at $\tau =0$ are
\begin{equation}
n_{\pm }(x,t,0)=\int_{0}^{t^{+}}\gamma (\tau )n_{\mp }(x,t,\tau )d\tau .
\label{initial0}
\end{equation}%
The main advantage of the system (\ref{str}) and (\ref{str2}) together with (%
\ref{initial}) and (\ref{initial0}) is that it is Markovian one. From this
system one can obtain various non-Markovian models including subdiffusive
and superdiffusive fractional equations. It can be done by eliminating the
residence time variable $\tau $ as in (\ref{den}) and introducing particular
models for the switching rate $\gamma (\tau )$.

\subsection{ Switching rate $\protect\gamma (\protect\tau )$}

One of the main purposes of this paper is to explore the anomalous case when
the switching rate $\gamma (\tau )$ is inversely proportional to the
residence time $\tau $ (negative aging). This rate describes the anomalous
persistence of a random walk: the longer a cell moves in a particular
direction without switching, the smaller the probability of switching to
another direction becomes. Here we consider two cases involving the
Mittag-Leffler function and Pareto distribution.

\textit{Case 1.} We make use of the following switching rate \cite{Cox}
\begin{equation}
\gamma (\tau )=-\frac{\dot{\Psi}\left( \tau \right) }{\Psi \left( \tau
\right) }  \label{ss}
\end{equation}%
with the survival probability \cite{Scalas}%
\begin{equation}
\Psi \left( \tau \right) =E_{\mu }\left[ -\left( \frac{\tau }{\tau _{0}}%
\right) ^{\mu }\right] ,\ 0<\mu <1,  \label{SSS}
\end{equation}%
where $\tau _{0}$ is the time constant, $E_{\mu }\left[ z\right] $ is the
Mittag-Leffler function.

\textit{Case 2. }We employ the explicit expression for the switching rate as
\cite{Fer,Steve}\textit{\ }
\begin{equation}
\gamma (\tau )=\frac{\mu }{\tau _{0}+\tau },\ 0<\mu <2.  \label{inverse}
\end{equation}%
This assumption together with (\ref{ss}) leads to a survival function $\Psi
(\tau )$ that has a power law dependence (Pareto distribution)
\begin{equation}
\Psi (\tau )=\left[ \frac{\tau _{0}}{\tau _{0}+\tau }\right] ^{\mu }.
\label{Pareto}
\end{equation}%
Our next step is to obtain the non-Markovian equations for $\rho _{+}(x,t)$
and $\rho _{-}(x,t)$ by eliminating the residence time variable $\tau $ (see
(\ref{den})).

\section{Non-Markovian master equations for $\protect\rho _{+}(x,t)$ and $%
\protect\rho _{-}(x,t)$}

The aim now is to find equations for $\rho _{+}(x,t)$ and $\rho _{-}(x,t)$
by solving the partial differential equations (\ref{str}) and (\ref{str2})
together with the boundary condition (\ref{initial0}) at\ $\tau =0$ and
initial condition (\ref{initial}) at $t=0$. By using the method of
characteristics we find for $\tau <t$
\begin{equation}
n_{\pm }(x,t,\tau )=n_{\pm }(x\mp v\tau ,t-\tau ,0)e^{-\int_{0}^{\tau
}\gamma (u)du}e^{-\theta \tau }.  \label{solution}
\end{equation}%
It is convenient to use the survival function from (\ref{ss})
\begin{equation}
\Psi (\tau )=e^{-\int_{0}^{\tau }\gamma (u)du}.  \label{Sur}
\end{equation}%
and the fluxes between two states (switching terms) $i_{+}(x,t)$ and $%
i_{-}(x,t):$
\begin{equation}
i_{\pm }(x,t)=\int_{0}^{t^{+}}\gamma (\tau )n_{\pm }(x,t,\tau )d\tau .
\label{i}
\end{equation}%
We notice that \ $n_{+}(x,t,0)=i_{-}(x,t)$ and $n_{-}(x,t,0)=i_{+}(x,t),$ so
the formula (\ref{solution}) can be rewritten as
\begin{equation}
n_{\pm }(x,t,\tau )=i_{\mp }(x\mp v\tau ,t-\tau )\Psi (\tau )e^{-\theta \tau
}.  \label{meaning}
\end{equation}%
This formula has a very simple meaning. For example, the density $%
n_{+}(x,t,\tau )$ gives the number of cells at point $x$ and time $t$ moving
in the right direction during time $\tau $ as a result of the following
process. The first factor in the RHS of (\ref{meaning}), $i_{-}(x-v\tau
,t-\tau ),$ gives the number of cells that switch their velocity from $-v$
to $v$ at the point $\ x-v\tau $ at the time $t-\tau $ and survive during
movement time $\tau $ due to random switching described by $\Psi (\tau )$
and the death process described by $e^{-\theta \tau }.$

The balance equations for the unstructured density $\rho _{\pm
}(x,t)=\int_{0}^{t^{+}}n_{+}(x,t,\tau )d\tau $ can be found by
differentiating (\ref{den}) together with (\ref{meaning}) with respect to
time $t$ or by using the Fourier-Laplace transform technique (see Appendix
1, part (B)). We obtain
\begin{equation}
\frac{\partial \rho _{+}}{\partial t}+v\frac{\partial \rho _{+}}{\partial x}%
=-i_{+}(x,t)+i_{-}(x,t)-\theta \rho _{+},  \label{mean3}
\end{equation}%
\begin{equation}
\frac{\partial \rho _{-}}{\partial t}-v\frac{\partial \rho _{-}}{\partial x}%
=i_{+}(x,t)-i_{-}(x,t)-\theta \rho _{-}.  \label{mean4}
\end{equation}%
These two equations have a similar structure to the standard model for a
persistent random walk with reactions \cite{MFH,Had,H,Hi,Bo}, but the
switching terms $i_{+}(x,t)$ and $i_{-}(x,t)$ are essentially different from
the simple Markovian terms $\gamma \rho _{+}$ and $\gamma \rho _{-}:$%
\begin{equation}
i_{+}(x,t)=\int_{0}^{t}K(t-\tau )\rho _{+}(x-v(t-\tau ),\tau )e^{-\theta
(t-\tau )}d\tau ,  \label{rate_i}
\end{equation}%
\begin{equation}
i_{-}(x,t)=\int_{0}^{t}K(t-\tau )\rho _{-}(x+v(t-\tau ),\tau )e^{-\theta
(t-\tau )}d\tau .  \label{rate_ii}
\end{equation}%
Here $K(\tau )$ is the memory kernel determined by its Laplace transform
\cite{Ke}
\begin{equation}
\hat{K}(s)=\frac{\hat{\psi}(s)}{\hat{\Psi}(s)},  \label{Kd}
\end{equation}%
where $\hat{\psi}(s)$ and $\hat{\Psi}(s)$ are the Laplace transforms of the
residence time density $\psi (\tau )=-d\Psi /d\tau $ and\ the survival
function $\Psi (\tau ).$ One can see that $i_{+}(x,t)$ and $i_{-}(x,t)$
depend on the death rate $\theta $ and transport process involving velocity $%
v.$ This is a non-Markovian effect \cite{Fedo1,Fedo11,Fedo2}. To obtain (\ref%
{rate_i}) and (\ref{rate_ii}), we use the Fourier-Laplace transform
\begin{equation}
\tilde{\imath}_{\pm }(k,s)=\int_{\mathbb{R}}\int_{0}^{t}i_{\pm
}(x,t)e^{ikx-st}dtdx,  \label{LF1}
\end{equation}%
\begin{equation}
\tilde{\rho}_{\pm }(k,s)=\int_{\mathbb{R}}\int_{0}^{t}\rho _{\pm
}(x,t)e^{ikx-st}dtdx.  \label{LF2}
\end{equation}%
We find (see Appendix 1, part (A))
\begin{equation}
\tilde{\imath}_{\pm }(k,s)=\frac{\hat{\psi}(s\mp ikv+\theta )}{\hat{\Psi}%
(s\mp ikv+\theta )}\tilde{\rho}_{\pm }(k,s).  \label{LLL}
\end{equation}%
Inverse Fourier-Laplace transform gives the explicit expressions for the
switching terms $i_{+}(x,t)$ and $i_{-}(x,t)$ in terms of the unstructured
densities $\rho _{+}(x,t)$ and $\rho _{-}(x,t).$

If we introduce the notations
\begin{equation*}
\hat{\Psi}_{\theta }^{\pm }=\hat{\Psi}(s\pm ikv+\theta ),\ \hat{\psi}%
_{\theta }^{\pm }=\hat{\psi}(s\pm ikv+\theta ),
\end{equation*}%
then the Fourier-Laplace transform of the total density $\rho (x,t)=\rho
_{+}(x,t)+\rho _{-}(x,t)$ can be written as (see Appendix 1, part (C))
\begin{equation}
\tilde{\rho}(k,s)=\frac{\rho _{+}^{0}(k)\left[ \hat{\Psi}_{\theta }^{-}+\hat{%
\Psi}_{\theta }^{+}\hat{\psi}_{\theta }^{-}\right] +\rho _{-}^{0}(k)\left[
\hat{\Psi}_{\theta }^{+}+\hat{\Psi}_{\theta }^{-}\hat{\psi}_{\theta }^{+}%
\right] }{1-\hat{\psi}_{\theta }^{+}\hat{\psi}_{\theta }^{-}},  \label{full}
\end{equation}%
where $\rho _{\pm }^{0}(k)=\int_{\mathbb{R}}\rho _{\pm }^{0}(x)e^{ikx}dx.$

\subsection{ Markovian two-state model}

If the switching rate $\gamma (\tau )$ is constant, it corresponds to the
exponential survival function $\Psi (\tau )=e^{-\gamma \tau }$ for which $%
\hat{K}(s)=\gamma $ and $K(\tau )=\gamma \delta \left( \tau \right) $. In
this case (\ref{mean3}) and (\ref{mean4}) can be reduced to a classical
two-state Markovian model for the density of cells moving right, $\rho
_{+}(x,t),$ and the density of cells moving left, $\rho _{-}(x,t):$
\begin{equation}
\frac{\partial \rho _{+}}{\partial t}+v\frac{\partial \rho _{+}}{\partial x}%
=-\gamma \left( \rho _{+}-\rho _{-}\right) -\theta \rho _{+},  \label{mean1}
\end{equation}%
\begin{equation}
\frac{\partial \rho _{-}}{\partial t}-v\frac{\partial \rho _{-}}{\partial x}%
=-\gamma \left( \rho _{+}-\rho _{-}\right) -\theta \rho _{-}.  \label{mean2}
\end{equation}%
When $\theta =0,$ the model is well known as the persistent random walk or
correlated random walk which was analyzed in \cite{Go,Kac}. The whole idea
of this random walk model was to remedy the unphysical property of Brownian
motion of infinite propagation. Two equations (\ref{mean1}) and (\ref{mean2}%
) can be rewritten as a telegraph equation for the total density $\rho
(x,t)=\rho _{-}(x,t)+\rho _{+}(x,t)$. This model covers the ballistic motion
and the standard diffusive motion in the limit $v\rightarrow \infty $ and $%
\gamma \rightarrow \infty $ such that $v^{2}/\gamma $ remains constant. The
Markovian model has been studied thoroughly and all details can be found in
\cite{MFH,Had,H,Hi,Bo}. We should mention that relatively simple extension
of the two-state Markovian dynamical system (\ref{mean1}) and (\ref{mean2})
is the non-Markovian model with the waiting time PDF of the form
\begin{equation*}
\psi \left( \tau \right) =\beta ^{2}\tau e^{-\beta \tau }.
\end{equation*}%
In this case, the Laplace transforms are
\begin{equation*}
\hat{\psi}(s)=\frac{\beta ^{2}}{\left( \beta +s\right) ^{2}},\quad \hat{K}%
(s)=\frac{s\hat{\psi}(s)}{1-\hat{\psi}(s)}=\frac{\beta ^{2}}{2\beta +s}.
\end{equation*}%
The memory kernel in (\ref{rate_i}) and (\ref{rate_ii}) has an exponential
form
\begin{equation*}
K\left( \tau \right) =\beta ^{2}e^{-2\beta \tau }.
\end{equation*}%
Non-Markovian random motions of particles with velocities alternating at
Erlang-distributed and gamma-distributed random times have been considered
in \cite{Ital0,Ital}. In this paper we will focus on the anomalous case
involving cells velocities alternating at power-law distributed random times
\cite{Mas,Fer,Sok}.

\subsection{Non-Markovian model involving anomalous switching}

Let us consider two anomalous cases when the switching rate $\gamma (\tau )$
(\ref{ss}) is inversely proportional to the residence time $\tau $.

\textit{Case 1.} The Laplace transforms of the survival function $\Psi
\left( \tau \right) =E_{\mu }\left[ -\left( \frac{\tau }{\tau _{0}}\right)
^{\mu }\right] \ $and $\psi (\tau )=-d\Psi \left( \tau \right) /d\tau $ are
\begin{equation}
\hat{\Psi}\left( s\right) =\frac{\tau _{0}^{\mu }s^{\mu -1}}{1+\left( s\tau
_{0}\right) ^{\mu }},\quad \ \hat{\psi}\left( s\right) =\frac{1}{1+\left(
s\tau _{0}\right) ^{\mu }}.
\end{equation}%
The Laplace transform of the memory kernel $K\left( \tau \right) $ is%
\begin{equation}
\hat{K}\left( s\right) =\frac{s^{1-\mu }}{\tau _{0}^{\mu }}.  \label{KK}
\end{equation}

\textit{Case 2.} The survival function $\Psi (\tau )$ has a Pareto
distribution (\ref{Pareto}) and corresponding waiting time PDF $\psi (\tau )$
is
\begin{equation}
\psi (\tau )=\frac{\mu \tau _{0}^{\mu }}{(\tau _{0}+\tau )^{1+\mu }}.
\label{eq:psi_tails}
\end{equation}%
When $0<\mu <1$, the asymptotic approximation for the Laplace transform $%
\hat{\psi}\left( s\right) $\ can be found from the Tauberian theorem \cite%
{Feller}
\begin{equation}
\hat{\psi}\left( s\right) \simeq 1-\Gamma (1-\mu )\tau _{0}{}^{\mu }s^{\mu
},\qquad s\rightarrow 0.  \label{anom}
\end{equation}%
The Laplace transform of memory kernel $K(\tau )$ can be written
approximately as
\begin{equation}
\hat{K}\left( s\right) \simeq \frac{s^{1-\mu }}{\Gamma (1-\mu )\tau
_{0}^{\mu }}.  \label{KKK}
\end{equation}%
Note that the only difference between (\ref{KK}) (case 1) and (\ref{KKK})
(case 2) is the $\Gamma (1-\mu )$ in the denominator in (\ref{KKK}).

\subsection{ Tempered fractional material derivatives}

In the anomalous case the switching terms (\ref{rate_i}) and (\ref{rate_ii})
can be written in terms of tempered fractional material derivatives. Using (%
\ref{LLL}) and (\ref{KK}) we write the Fourier-Laplace transforms of $%
i_{+}(x,t)$ and $i_{-}(x,t)$ as
\begin{equation}
\tilde{\imath}_{\pm }(k,s)=\tau _{0}^{-\mu }(s\mp ikv+\theta )^{1-\mu }%
\tilde{\rho}_{\pm }(k,s).  \label{if}
\end{equation}%
We define the tempered fractional material derivatives $\left( \frac{%
\partial }{\partial t}\pm v\frac{\partial }{\partial x}+\theta \right)
^{1-\mu }$ of order $1-\mu $ by their Fourier-Laplace transforms
\begin{equation}
\mathcal{LF}\left\{ \left( \frac{\partial }{\partial t}\pm v\frac{\partial }{%
\partial x}+\theta \right) ^{1-\mu }\rho \right\} =(s\pm ikv+\theta )^{1-\mu
}\tilde{\rho},\quad 0<\mu <1.  \label{op}
\end{equation}%
Note that fractional material derivatives with the factor $(s\pm ik)^{1-\mu
} $ have been introduced in \cite{Sok}. Evolution equations for anomalous
diffusion involving coupled space-time\ fractional derivative operators
involving the Fourier-Laplace symbols like $(s+ik)^{\beta },$ $%
(s+k^{2})^{\beta },$ etc. have been considered in \cite{Mer0,Mer1,Mer3}.
Here we have the tempered fractional derivative operator (\ref{op}) that
involves both the advective transport \ and the death rate $\theta .$ The
latter plays the role of tempering parameter because $(s\pm ikv+\theta
)^{1-\mu }$ has a finite limit $\theta ^{1-\mu }$ as $s\rightarrow 0$ and $%
k\rightarrow 0$. We represent the anomalous switching terms as%
\begin{equation*}
i_{\pm }(x,t)=\tau _{0}^{-\mu }\left( \frac{\partial }{\partial t}\mp v\frac{%
\partial }{\partial x}+\theta \right) ^{1-\mu }\rho _{\pm },\quad 0<\mu <1.
\end{equation*}%
The master equations (\ref{mean3}) and (\ref{mean4}) can be rewritten as
\begin{equation}
\frac{\partial \rho _{+}}{\partial t}+v\frac{\partial \rho _{+}}{\partial x}%
=-\tau _{0}^{-\mu }\left( \frac{\partial }{\partial t}-v\frac{\partial }{%
\partial x}+\theta \right) ^{1-\mu }\rho _{+}+\tau _{0}^{-\mu }\left( \frac{%
\partial }{\partial t}+v\frac{\partial }{\partial x}+\theta \right) ^{1-\mu
}\rho _{-}-\theta \rho _{+},  \label{anom1}
\end{equation}%
\begin{equation}
\frac{\partial \rho _{-}}{\partial t}-v\frac{\partial \rho _{-}}{\partial x}%
=-\tau _{0}^{-\mu }\left( \frac{\partial }{\partial t}+v\frac{\partial }{%
\partial x}+\theta \right) ^{1-\mu }\rho _{-}+\tau _{0}^{-\mu }\left( \frac{%
\partial }{\partial t}-v\frac{\partial }{\partial x}+\theta \right) ^{1-\mu
}\rho _{+}-\theta \rho _{-}.  \label{anom2}
\end{equation}%
Note that when $\theta =0$ these equations describe a very strong
persistence in a particular direction. For the symmetrical initial
conditions
\begin{equation*}
\rho _{+}^{0}(x)=\frac{1}{2}\delta \left( x\right) ,\quad \rho _{-}^{0}(x)=%
\frac{1}{2}\delta \left( x\right)
\end{equation*}%
for which $\mathbb{E}\left\{ x(t)\right\} =0$, the mean squared displacement
$\mathbb{E}\left\{ x^{2}(t)\right\} $ exhibits ballistic behaviour \cite%
{Mas,Fer,Sok}:%
\begin{equation*}
\mathbb{E}\left\{ x^{2}(t)\right\} \simeq t^{2}.
\end{equation*}%
However, if all cells at $t=0$ start to move to the right with the velocity $%
v$ from the point $x=0:$
\begin{equation*}
\rho _{+}^{0}(x)=\delta \left( x\right) ,\quad \rho _{-}^{0}(x)=0,
\end{equation*}%
then (see Appendix 2) the first moment $\mathbb{E}\left\{ x(t)\right\} $ is
\begin{equation*}
\mathbb{E}\left\{ x(t)\right\} \simeq \frac{v\tau _{0}^{\mu }}{2}t^{1-\mu }.
\end{equation*}%
The subballistic behaviour of $\mathbb{E}\left\{ x(t)\right\} $ was obtained
in \cite{Barkai} for the fractional Kramers equation.

In the large scale limit $k\rightarrow 0,$ we expand $(s+\theta +ikv)^{1-\mu
}=\allowbreak \left( s+\theta \right) ^{1-\mu }+ikv\left( 1-\mu \right)
\left( s+\theta \right) ^{-\mu }+o(k)$ and obtain from (\ref{if})
\begin{equation}
\tilde{\imath}_{+}(k,s)=\tau _{0}^{-\mu }\left[ (s+\theta )^{1-\mu
}-ikv\left( 1-\mu \right) \left( s+\theta \right) ^{-\mu }\right] \tilde{\rho%
}_{+},
\end{equation}%
\begin{equation}
\tilde{\imath}_{-}(k,s)=\tau _{0}^{-\mu }\left[ (s+\theta )^{1-\mu
}+ikv\left( 1-\mu \right) \left( s+\theta \right) ^{-\mu }\right] \tilde{\rho%
}_{-}.
\end{equation}%
By using inverse the Fourier-Laplace transform we find\
\begin{eqnarray}
i_{+}(x,t) &=&e^{-\theta t}\frac{\partial }{\partial t}\int_{0}^{t}m_{\mu
}(t-\tau )\rho _{+}(x,\tau )e^{\theta \tau }d\tau   \notag \\
&&-\left( 1-\mu \right) e^{-\theta t}v\int_{0}^{t}m_{\mu }(t-\tau )\frac{%
\partial \rho _{+}(x,\tau )}{\partial x}e^{\theta \tau }d\tau ,
\end{eqnarray}%
\begin{eqnarray}
i_{-}(x,t) &=&e^{-\theta t}\frac{\partial }{\partial t}\int_{0}^{t}m_{\mu
}(t-\tau )\rho _{-}(x,\tau )e^{\theta \tau }d\tau +  \notag \\
&&+\left( 1-\mu \right) e^{-\theta t}v\int_{0}^{t}m_{\mu }(t-\tau )\frac{%
\partial \rho _{-}(x,\tau )}{\partial x}e^{\theta \tau }d\tau ,
\end{eqnarray}%
where $m_{\mu }(t)$ is the classical renewal measure density associated with
the survival probability (\ref{SSS})
\begin{equation}
m_{\mu }(t)=\frac{t^{\mu -1}}{\Gamma \left( \mu \right) \tau _{0}^{\mu }}%
,\quad 0<\mu <1.
\end{equation}%
The density $m_{\mu }(t)$ has a meaning of the average number of jumps per
unit time. Note that the switching terms $i_{+}(x,t)$ and $i_{-}(x,t)$
involves the advection term with memory effects. This coupling of advection
with switching rate is a pure non-Markovian effect. Expressions for $%
i_{+}(x,t)$ and $i_{-}(x,t)$ can be rewritten with the standard notations
involving the Riemann-Liouville fractional derivative $\mathcal{D}%
_{t}^{1-\mu }$of order $1-\mu $ and fractional integral $I_{t}^{\mu }$of
order $\mu $
\begin{equation}
i_{+}(x,t)=e^{-\theta t}\mathcal{D}_{t}^{1-\mu }\left[ \rho
_{+}(x,t)e^{\theta t}\right] -\left( 1-\mu \right) e^{-\theta t}vI_{t}^{\mu }%
\left[ \frac{\partial \rho _{+}(x,t)}{\partial x}e^{\theta t}\right] ,
\notag
\end{equation}%
\begin{equation}
i_{+}(x,t)=e^{-\theta t}\mathcal{D}_{t}^{1-\mu }\left[ \rho
_{-}(x,t)e^{\theta t}\right] +\left( 1-\mu \right) e^{-\theta t}vI_{t}^{\mu }%
\left[ \frac{\partial \rho _{-}(x,t)}{\partial x}e^{\theta t}\right] .
\notag
\end{equation}

It is easy to generalize the master equations (\ref{anom1}) and (\ref{anom2}%
) for the situation when the cells motility involves the random Brownian
motion with diffusion coefficient $D$. We can write
\begin{equation*}
\frac{\partial \rho _{+}}{\partial t}+v\frac{\partial \rho _{+}}{\partial x}%
=D\frac{\partial ^{2}\rho _{+}}{\partial x^{2}}-\tau _{0}^{-\mu }\left(
\mathcal{D}_{\theta }^{-}\rho _{+}-\mathcal{D}_{\theta }^{+}\rho _{-}\right)
-\theta \rho _{+},
\end{equation*}%
\begin{equation*}
\frac{\partial \rho _{-}}{\partial t}-v\frac{\partial \rho _{-}}{\partial x}%
=D\frac{\partial ^{2}\rho _{-}}{\partial x^{2}}-\tau _{0}^{-\mu }\left(
\mathcal{D}_{\theta }^{+}\rho _{-}-\mathcal{D}_{\theta }^{-}\rho _{+}\right)
-\theta \rho _{-},
\end{equation*}%
where the tempered fractional derivatives $\mathcal{D}_{\theta }^{\pm }\rho $
are defined by
\begin{equation*}
\mathcal{LF}\left\{ \mathcal{D}_{\theta }^{\pm }\rho \right\} =(s\pm
ikv+\theta -Dk^{2})^{1-\mu }\tilde{\rho}.
\end{equation*}

\subsection{ Tempered superdiffusion}

Now let us find the switching terms (\ref{rate_i}) and (\ref{rate_ii}) in
the case when the first moment $<T>=\int_{0}^{\infty }\tau \psi (\tau )\tau $
is finite, while the variance is divergent $1<\mu <2$. When the death rate $%
\theta =0,$ mean squared displacement $\mathbb{E}\left\{ x^{2}(t)\right\} $
exhibits subballistic superdiffusive behaviour \cite{Mas,Fer,Sok}%
\begin{equation*}
\mathbb{E}\left\{ x^{2}(t)\right\} \simeq t^{3-\mu }
\end{equation*}%
\ (see Appendix 3). In this case the small $s$ expansion of $\hat{\psi}%
\left( s\right) $ gives
\begin{equation}
\hat{\psi}\left( s\right) \simeq 1-<T>s+A<T>s^{\mu },\qquad 1<\mu <2.
\label{small2}
\end{equation}%
Then
\begin{equation*}
\hat{K}(s)=\frac{s\hat{\psi}(s)}{1-\hat{\psi}(s)}\simeq \frac{1}{<T>}\left(
1+As^{\mu -1}\right) .
\end{equation*}%
Using (\ref{LLL}) and (\ref{KK}) we write the Fourier-Laplace transforms of $%
i_{+}(x,t)$ and $i_{-}(x,t)$ as
\begin{equation}
\tilde{\imath}_{\pm }(k,s)=\frac{1}{<T>}\left( 1+A\left( s+\theta \mp
ikv\right) ^{\mu -1}\right) \tilde{\rho}_{\pm }(k,s).  \label{sab}
\end{equation}%
One can introduce the tempered fractional material derivatives $\left( \frac{%
\partial }{\partial t}\pm v\frac{\partial }{\partial x}+\theta \right) ^{\mu
-1}$ of order $\mu -1$ for intermediate subballistic superdiffusive case $%
1<\mu <2$ as
\begin{equation}
\mathcal{LF}\left\{ \left( \frac{\partial }{\partial t}\pm v\frac{\partial }{%
\partial x}+\theta \right) ^{\mu -1}\rho \right\} =\left( s\pm ikv+\theta
\right) ^{\mu -1}\tilde{\rho},\quad 1<\mu <2.
\end{equation}%
The switching terms can be written as%
\begin{equation}
i_{\pm }(x,t)=\frac{1}{<T>}\left( 1+A\left( \frac{\partial }{\partial t}\mp v%
\frac{\partial }{\partial x}+\theta \right) ^{\mu -1}\right) \rho _{\pm }.
\end{equation}%
In the limit $k\rightarrow 0,$ we use the expansion $(s+ikv+\theta )^{\mu
-1}=\allowbreak \left( s+\theta \right) ^{\mu -1}+ikv\left( \mu -1\right)
\left( s+\theta \right) ^{\mu -2}+o(k)$ to obtain from (\ref{sab})
\begin{equation}
\tilde{\imath}_{+}(k,s)=\frac{1}{<T>}\left[ 1+A(s+\theta )^{\mu -1}-A\left(
s+\theta \right) ^{\mu -2}ikv\left( \mu -1\right) \right] \tilde{\rho}_{+},
\end{equation}%
\begin{equation}
\tilde{\imath}_{-}(k,s)=\frac{1}{<T>}\left[ 1+A(s+\theta )^{\mu -1}+A\left(
s+\theta \right) ^{\mu -2}ikv\left( \mu -1\right) \right] \tilde{\rho}_{-}.
\end{equation}%
By using inverse the Fourier-Laplace transform we find\
\begin{eqnarray}
i_{+}(x,t) &=&\frac{\rho _{+}(x,t)}{<T>}+e^{-\theta t}\frac{\partial }{%
\partial t}\int_{0}^{t}m_{A}(t-\tau )\rho _{+}(x,\tau )e^{\theta \tau }d\tau
\notag \\
&&-v\left( \mu -1\right) e^{-\theta t}\int_{0}^{t}m_{A}(t-\tau )\frac{%
\partial \rho _{+}(x,\tau )}{\partial x}e^{\theta \tau }d\tau ,
\end{eqnarray}%
\begin{eqnarray}
i_{-}(x,t) &=&\frac{\rho _{-}(x,t)}{<T>}+e^{-\theta t}\frac{\partial }{%
\partial t}\int_{0}^{t}m_{A}(t-\tau )\rho _{-}(x,\tau )e^{\theta \tau }d\tau
+  \notag \\
&&+v\left( \mu -1\right) e^{-\theta t}\int_{0}^{t}m_{A}(t-\tau )\frac{%
\partial \rho _{-}(x,\tau )}{\partial x}e^{\theta \tau }d\tau ,
\end{eqnarray}%
where
\begin{equation}
m_{A}(t)=\frac{At^{1-\mu }}{<T>\Gamma \left( 2-\mu \right) },\qquad 1<\mu <2.
\end{equation}%
Switching terms $i_{+}(x,t)$ and $i_{-}(x,t)$ can be rewritten in terms of
the Riemann-Liouville fractional derivative $\mathcal{D}_{t}^{\mu -1}$of
order $\mu -1$ and fractional integral $I_{t}^{2-\mu }$of order $2-\mu .$
Now we are in a position to discuss the implications of tempering due to the
random death process. In the next subsection we consider the stationary case.

\section{ Stationary profile and truncated L\'{e}vy flights.}

The aim of this section is to analyze the cell density profiles in the
stationary case for the strong anomalous case $0<\mu <1$. To ensure the
existence of stationary profiles $\rho _{+}^{s}(x)$ and $\rho _{-}^{s}(x)$,
we introduce the constant source of cells at the point $x=0$. We keep in
mind the problem of cancer cell proliferation. One can think of the tumor
consisting of the tumor core with a high density of cells (proliferation
zone) at $x=0$ and the outer invasive zone where the cell density is
smaller. We are interested in the stationary profile of cancer cells
spreading in the outer migrating zone \cite{Fedo1}. For simplicity we
consider only one-dimensional case here. The generalization for 2-D and 3-D
cases can be made in the standard way \cite{Fedo1}.

Let us find a stationary solution to the system (\ref{anom1}) and (\ref%
{anom2}). Now we show that in long-time limit master equations can be
written in terms of exponentially truncated fractional derivatives in which
the ratio $\theta /v$ plays the role of tempering to a L\'{e}vy jump
distribution. The profiles $\rho _{+}^{s}(x)$ and $\rho _{-}^{s}(x)$ can be
found from
\begin{equation}
v\frac{\partial \rho _{+}^{s}(x)}{\partial x}=-i_{+}^{s}(x)+i_{-}^{s}(x)-%
\theta \rho _{+}^{s}(x),  \label{st1}
\end{equation}%
\begin{equation}
-v\frac{\partial \rho _{-}^{s}(x)}{\partial x}=i_{+}^{s}(x)-i_{-}^{s}(x)-%
\theta \rho _{-}^{s}(x),  \label{st2}
\end{equation}%
where $i_{+}^{s}(x)$ and $i_{-}^{s}(x)$ are the stationary switching terms
with the Fourier transforms:
\begin{equation}
\tilde{\imath}_{+}^{s}(k)=\frac{\left( -ikv+\theta )\right) ^{1-\mu }}{\tau
_{0}^{\mu }}\tilde{\rho}_{+}^{s}(k),  \label{st3}
\end{equation}%
\begin{equation}
\tilde{\imath}_{-}^{s}(k)=\frac{\left( ikv+\theta )\right) ^{1-\mu }}{\tau
_{0}^{\mu }}\tilde{\rho}_{+}^{s}(k).  \label{st4}
\end{equation}%
These formulas are obtain from (\ref{if}) as $s\rightarrow 0$ ($t\rightarrow
\infty $). Using the shift theorem we can write $i_{+}^{s}(x)$ and $%
i_{-}^{s}(x)$ in terms of exponentially truncated fractional derivatives
\cite{Car}
\begin{equation}
i_{+}^{s}(x)=\frac{v^{1-\mu }e^{-\frac{\theta x}{v}}\left( _{-\infty }\emph{D%
}^{1-\mu }\left[ e^{\frac{\theta x}{v}}\rho _{+}^{s}(x)\right] \right) }{%
\tau _{0}^{\mu }},
\end{equation}%
\begin{equation}
i_{-}^{s}(x)=\frac{v^{1-\mu }e^{\frac{\theta x}{v}}\left( \emph{D}_{\infty
}^{1-\mu }\left[ e^{-\frac{\theta x}{v}}\rho _{-}^{s}(x)\right] \right) }{%
\tau _{0}^{\mu }}.
\end{equation}%
Here $_{-\infty }\emph{D}^{1-\mu }$ and $\emph{D}_{\infty }^{1-\mu }$ are
the Weyl derivatives of order $1-\mu $ \cite{Samko}
\begin{equation}
_{-\infty }\emph{D}^{1-\mu }\rho (x)=\frac{1}{\Gamma (\mu )}\frac{d}{dx}%
\int_{-\infty }^{x}\frac{\rho (y)dy}{(x-y)^{1-\mu }},
\end{equation}%
\begin{equation}
\emph{D}_{\infty }^{1-\mu }\rho (x)=-\frac{1}{\Gamma (\mu )}\frac{d}{dx}%
\int_{x}^{\infty }\frac{\rho (y)dy}{(y-x)^{1-\mu }}
\end{equation}%
with the Fourier transforms
\begin{equation*}
\mathcal{F}\left\{ _{-\infty }\emph{D}^{1-\mu }\rho (x)\right\} =\left(
-ik\right) ^{1-\mu }\hat{\rho}(k)
\end{equation*}%
and
\begin{equation*}
\mathcal{F}\left\{ \emph{D}_{\infty }^{1-\mu }\rho (x)\right\} =\left(
ik\right) ^{1-\mu }\hat{\rho}(k).
\end{equation*}%
We should note that our theory with death rate tempering is fundamentally
different from the standard tempering \cite{Car,Ba,Mer}, which is just the
truncation of the power law jump distribution by an exponential factor
involving a tempering parameter. In fact we do\ not introduce the L\'{e}vy
jump distribution functions at all. It means that we are not just employing
a mathematical trick to overcome long jumps with infinite variance which is
a standard problem of L\'{e}vy flights.

\subsection{ Upper and lower bounds for the stationary profiles}

The purpose of this subsection is to find the upper bound, $\rho _{u}(x),$
and\ the lower bound, $\rho _{l}(x),$ for the stationary profile $\rho
^{s}(x)=\rho _{+}^{s}(x)+\rho _{-}^{s}(x)$ in the strong anomalous case $\mu
<1:$
\begin{equation*}
\rho ^{l}(x)<\rho ^{s}(x)<\rho ^{u}(x).
\end{equation*}%
If cells are released at the point $x=0$ at the constant rate $g$ on the
right and at the same rate $g$ on the left, then the upper bound can be
easily found from the advection-reaction equation
\begin{equation*}
v\frac{\partial \rho ^{u}(x)}{\partial x}=-\theta \rho ^{u}(x).
\end{equation*}%
Clearly this equation describes the ballistic motion of cells without
switching. We obtain
\begin{equation}
\rho ^{u}(x)=\frac{g}{v}\exp \left[ -\frac{\theta |x|}{v}\right] ,
\label{Upper}
\end{equation}%
where the prefactor $g/v$ is found from the condition $g=\theta
\int_{0}^{\infty }\rho ^{u}(x)dx$ \cite{Steve}.

We can find the\ lower bound $\rho ^{l}(x)$ using the small $k$ expansion
\begin{equation}
(\theta \pm ikv)^{1-\mu }=\theta ^{1-\mu }\allowbreak \pm ikv\theta ^{-\mu
}\left( 1-\mu \right) +\allowbreak O\left( k^{2}\right) .  \label{expa}
\end{equation}%
From (\ref{st3}) and (\ref{st4}) we get
\begin{equation*}
\tilde{\imath}_{+}^{s}(k)=\frac{\theta }{\left( \theta \tau _{0}\right)
^{\mu }}\tilde{\rho}_{+}^{s}(k)-\frac{ikv\left( 1-\mu \right) }{\left(
\theta \tau _{0}\right) ^{\mu }}\tilde{\rho}_{+}^{s}(k),
\end{equation*}%
\begin{equation*}
\tilde{\imath}_{-}^{s}(k)=\frac{\theta }{\left( \theta \tau _{0}\right)
^{\mu }}\tilde{\rho}_{-}^{s}(k)+\frac{ikv\left( 1-\mu \right) }{\left(
\theta \tau _{0}\right) ^{\mu }}\tilde{\rho}_{-}^{s}(k).
\end{equation*}%
Inverse Fourier transform gives%
\begin{equation}
i_{+}^{s}(x)=\frac{\theta }{\left( \theta \tau _{0}\right) ^{\mu }}\rho
_{+}^{s}(x)-\frac{v\left( 1-\mu \right) }{\left( \theta \tau _{0}\right)
^{\mu }}\frac{\partial \rho _{+}^{s}(x)}{\partial x},  \label{ist1}
\end{equation}%
\begin{equation}
i_{-}^{s}(x)=\frac{\theta }{\left( \theta \tau _{0}\right) ^{\mu }}\rho
_{-}^{s}(x)+\frac{v\left( 1-\mu \right) }{\left( \theta \tau _{0}\right)
^{\mu }}\frac{\partial \rho _{-}^{s}(x)}{\partial x}.  \label{ist2}
\end{equation}%
Note that the stationary switching terms $i_{+}^{s}(x)$ and $i_{-}^{s}(x)$
involve the advection terms proportional to the gradient of density. This is
a non-Markovian effect. Obviously advection terms are zero when $\mu =1.$
Under the condition of a weak death rate $\tau _{0}\theta <<1,$ we obtain
from (\ref{st1}), (\ref{st2}) together with (\ref{ist1}), (\ref{ist2}) the
following equation for $\rho ^{s}(x)=\rho _{+}^{s}(x)+\rho _{-}^{s}(x):$
\begin{equation}
D\frac{\partial ^{2}\rho ^{s}(x)}{\partial x^{2}}-\theta \rho ^{s}(x)=0,
\label{dii}
\end{equation}%
where $D$ is the effective diffusion coefficient
\begin{equation}
D=\frac{v^{2}}{\theta }\left( 1-\mu \right) ,\quad \mu <1.
\end{equation}%
Note that the diffusion coefficient $D$ depends on the death rate $\theta .$
The solution to (\ref{dii}) gives the lower bound
\begin{equation}
\rho ^{l}(x)=\frac{g}{v\sqrt{\left( 1-\mu \right) }}\exp \left[ -\frac{%
\theta |x|}{v\sqrt{\left( 1-\mu \right) }}\right] .  \label{Low}
\end{equation}%
Monte Carlo simulations involving $N=1000$ particles up to time $t=10^{3}$
confirm this bound$.$ One can see from Fig. 1 that apart from the very long
distance $\sim 10^{3},$ the Monte Carlo profile (black line) lies between
the upper bound (\ref{Upper}) (blue line) and the lower bound (\ref{Low})
(red line). Green line represents the best fit.

\begin{figure}[tbp]
\includegraphics[scale=0.59]{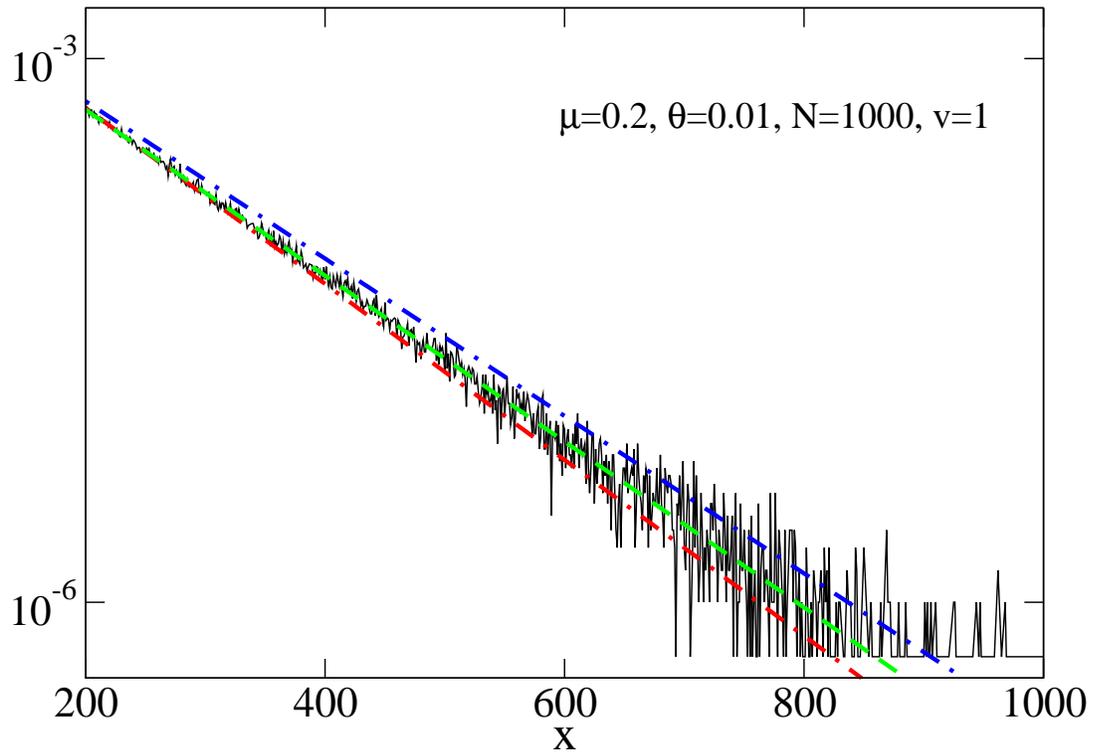}
\caption{Monte Carlo profile involving $N=10^3$ cells at time $t=10^3$
(black line), stationary upper bound (\protect\ref{Upper}) (blue line) and
lower bound (\protect\ref{Low}) (red line) profiles with parameters: the
anomalous exponent $\protect\mu =0.2,$ the death rate $\protect\theta =0.01,$
the time unit $\protect\tau_{0} =1$ and the cell velocity $v=1.$}
\label{fig:UpperBound}
\end{figure}

\section{ Conclusion}

We have been motivated by experiments showing non-Markovian subballistic
superdiffusive dynamics of cells \cite{J,PK,Jap, Mierke,Mierke2}. The main
challenge was to implement the random death process into a non-Markovian
transport processes governed by the anomalously persistent random walks. We
presented a Markovian model of cell motility that accounts for the effects
of a random death process and the dependence of switching rates on the
residence time variable $\tau $. Our purpose was to extend the the standard
model for the velocity-jump random walk with reactions for the anomalous
case of L\'{e}vy walks involving intermediate subballistic superdiffusive
motion. We derived non-Markovian master equations for the cell densities
with the generalized switching terms involving the tempered fractional
material derivatives. The cell degradation rate plays the role of a
tempering parameter. In the long-time limit we derived stationary master
equations in terms of exponentially truncated fractional derivatives in
which the rate of death tempers a L\'{e}vy jump distribution. We find the
upper and \ lower bounds for the stationary profiles corresponding to the
ballistic transport and diffusion with the death rate dependent diffusion
coefficient. Monte Carlo simulations confirm these bounds.

The main advantage of our model is that it can be extended to the case of
nonlinear death rate $\theta (\rho )$ that depends on the total density of
cells $\rho $. An important application of the results of this paper may be
the problem of wave propagation in reaction--transport systems involving
random walks with finite jump speed and memory effects \cite{Men0,Men}. It
would also be interesting to explore the long-memory effects in the context
of persistent random walks with random velocities \cite{Z}. It is also of
great interest to analyze the nonlinear tempering phenomenon leading to the
nonlinear diffusion \cite{St}.

\section{Acknowledgment.}

This work was funded by EPSRC grant EP/J019526/1. The authors wish to thank
Nickolay Korabel and Steven Falconer for very useful discussions.

\section{Appendix 1}

The purposes of this Appendix are (A) to express the switching functions $%
i_{+}(x,t)$ and $i_{-}(x,t)$ in terms of $\rho _{+}(x,t)$ and $\rho
_{-}(x,t);$ (B) to derive the master equations for the unstructured density $%
\rho _{+}(x,t),$ $\rho _{-}(x,t)$ (\ref{mean3}) and (\ref{mean4}); (C) to
find the Fourier-Laplace transform of the total density $\rho (x,t)=\rho
_{+}(x,t)+\rho _{-}(x,t).$

(A) Substitution of (\ref{meaning}) into (\ref{den}) and (\ref{i}) together
with the initial condition (\ref{initial}) gives
\begin{equation*}
i_{+}(x,t)=\int_{0}^{t^{-}}i_{-}(x-v\tau ,t-\tau )\psi (\tau )e^{-\theta
\tau }d\tau +\rho _{+}^{0}(x-vt)\psi (t)e^{-\theta t},
\end{equation*}%
\begin{equation}
i_{-}(x,t)=\int_{0}^{t^{-}}i_{+}(x+v\tau ,t-\tau )\psi (\tau )e^{-\theta
\tau }d\tau +\rho _{-}^{0}(x+vt)\psi (t)e^{-\theta t}  \label{int2}
\end{equation}%
and%
\begin{equation*}
\rho _{+}(x,t)=\int_{0}^{t^{-}}i_{-}(x-v\tau ,t-\tau )\Psi \left( \tau
\right) e^{-\theta \tau }d\tau +\rho _{+}^{0}(x-v\tau )\Psi \left( t\right)
e^{-\theta t},
\end{equation*}%
\begin{equation}
\rho _{-}(x,t)=\int_{0}^{t^{-}}i_{+}(x+v\tau ,t-\tau )\Psi \left( \tau
\right) e^{-\theta \tau }d\tau +\rho _{-}^{0}(x+v\tau )\Psi \left( t\right)
e^{-\theta t}.  \label{int3}
\end{equation}%
Applying the Fourier-Laplace transform together with shift theorem to above
equations, we find expressions for $i_{+}(x,t)$ and $i_{-}(x,t)$ in terms of
$\rho _{+}(x,t)$ and $\rho _{-}(x,t).$ By using (\ref{LF1}) and (\ref{LF2}),
we obtain from (\ref{int2}) and (\ref{int3})
\begin{equation}
\tilde{\imath}_{\pm }(k,s)=\left[ \tilde{\imath}_{\mp }(k,s)+\rho _{\pm
}^{0}(k)\right] \hat{\psi}(s\mp ikv+\theta ),  \label{pp}
\end{equation}%
\begin{equation}
\tilde{\rho}_{\pm }(k,s)=\left[ \tilde{\imath}_{\mp }(k,s)+\rho _{\pm
}^{0}(k)\right] \hat{\Psi}(s\mp ikv+\theta ).  \label{ii}
\end{equation}%
Therefore
\begin{equation*}
\tilde{\imath}_{\pm }(k,s)=\frac{\hat{\psi}(s\mp ikv+\theta )}{\hat{\Psi}%
(s\mp ikv+\theta )}\tilde{\rho}_{\pm }(k,s).
\end{equation*}%
\bigskip The inverse Fourier-Laplace transform gives (\ref{rate_i}).

(B) It is convenient to introduce the following notations
\begin{equation*}
\hat{\Psi}_{\theta }^{\pm }=\hat{\Psi}(s\pm ikv+\theta ),\ \hat{\psi}%
_{\theta }^{\pm }=\hat{\psi}(s\pm ikv+\theta ),
\end{equation*}%
then solving (\ref{pp}) and (\ref{ii}) for $\tilde{\rho}_{+}$ and $\tilde{%
\rho}_{-}$ we find%
\begin{equation}
\tilde{\rho}_{+}(k,s)=\left[ \frac{\hat{\psi}_{\theta }^{+}\tilde{\rho}%
_{-}(k,s)}{\hat{\Psi}_{\theta }^{+}}+\rho _{+}^{0}(k)\right] \hat{\Psi}%
_{\theta }^{-},  \label{ro1}
\end{equation}%
\begin{equation}
\tilde{\rho}_{-}(k,s)=\left[ \frac{\hat{\psi}_{\theta }^{-}\tilde{\rho}%
_{+}(k,s)}{\hat{\Psi}_{\theta }^{-}}+\rho _{-}^{0}(k)\right] \hat{\Psi}%
_{\theta }^{+}.  \label{ro2}
\end{equation}%
These two equations can be rewritten as
\begin{equation*}
\frac{\tilde{\rho}_{+}(k,s)}{\hat{\Psi}_{\theta }^{-}}-\rho _{+}^{0}(k)=%
\frac{\hat{\psi}_{\theta }^{+}\tilde{\rho}_{-}(k,s)}{\hat{\Psi}_{\theta }^{+}%
},
\end{equation*}%
\begin{equation*}
\frac{\tilde{\rho}_{-}(k,s)}{\hat{\Psi}_{\theta }^{+}}-\rho _{-}^{0}(k)=%
\frac{\hat{\psi}_{\theta }^{-}\tilde{\rho}_{+}(k,s)}{\hat{\Psi}_{\theta }^{-}%
}.
\end{equation*}%
Then
\begin{equation*}
\frac{\tilde{\rho}_{+}(k,s)}{\hat{\Psi}_{\theta }^{-}}\left[ 1-\hat{\psi}%
_{\theta }^{-}\right] -\rho _{+}^{0}(k)=-\frac{\hat{\psi}_{\theta }^{-}%
\tilde{\rho}_{+}(k,s)}{\hat{\Psi}_{\theta }^{-}}+\frac{\hat{\psi}_{\theta
}^{+}\tilde{\rho}_{-}(k,s)}{\hat{\Psi}_{\theta }^{+}},
\end{equation*}%
\begin{equation*}
\frac{\tilde{\rho}_{-}(k,s)}{\hat{\Psi}_{\theta }^{+}}\left[ 1-\hat{\psi}%
_{\theta }^{+}\right] -\rho _{-}^{0}(k)=-\frac{\hat{\psi}_{\theta }^{+}%
\tilde{\rho}_{-}(k,s)}{\hat{\Psi}_{\theta }^{+}}+\frac{\hat{\psi}_{\theta
}^{-}\tilde{\rho}_{+}(k,s)}{\hat{\Psi}_{\theta }^{-}}.
\end{equation*}%
Since $\left[ 1-\hat{\psi}_{\theta }^{\pm }\right] /\hat{\Psi}_{\theta
}^{\pm }=s\mp ikv+\theta $, we obtain
\begin{equation*}
\left( s+ikv+\theta \right) \tilde{\rho}_{+}(k,s)-\rho _{+}^{0}(k)=-\frac{%
\hat{\psi}_{\theta }^{-}\tilde{\rho}_{+}(k,s)}{\hat{\Psi}_{\theta }^{-}}+%
\frac{\hat{\psi}_{\theta }^{+}\tilde{\rho}_{-}(k,s)}{\hat{\Psi}_{\theta }^{+}%
},
\end{equation*}%
\begin{equation*}
\left( s-ikv+\theta \right) \tilde{\rho}_{-}(k,s)-\rho _{-}^{0}(k)=-\frac{%
\hat{\psi}_{\theta }^{+}\tilde{\rho}_{-}(k,s)}{\hat{\Psi}_{\theta }^{+}}+%
\frac{\hat{\psi}_{\theta }^{-}\tilde{\rho}_{+}(k,s)}{\hat{\Psi}_{\theta }^{-}%
}.
\end{equation*}%
The left-hand sides are the Fourier-Laplace transforms of $\partial \rho
_{\pm }/\partial t\pm \partial \rho _{\pm }/\partial x-\theta \rho _{+},$
therefore, these two equations are the the Fourier-Laplace transforms of the
master equations (\ref{mean3}) and (\ref{mean4}).

(C) From (\ref{ro1}) and (\ref{ro2}) we find explicit expressions for $%
\tilde{\rho}_{+}(k,s)$ and $\tilde{\rho}_{-}(k,s):$
\begin{equation}
\tilde{\rho}_{+}(k,s)=\frac{\rho _{+}^{0}(k)\hat{\Psi}_{\theta }^{-}+\rho
_{-}^{0}(k)\hat{\Psi}_{\theta }^{-}\hat{\psi}_{\theta }^{+}}{1-\hat{\psi}%
_{\theta }^{+}\hat{\psi}_{\theta }^{-}},  \label{aa}
\end{equation}%
\begin{equation}
\tilde{\rho}_{-}(k,s)=\frac{\rho _{-}^{0}(k)\hat{\Psi}_{\theta }^{+}+\rho
_{+}^{0}(k)\hat{\Psi}_{\theta }^{+}\hat{\psi}_{\theta }^{-}}{1-\hat{\psi}%
_{\theta }^{+}\hat{\psi}_{\theta }^{-}}.  \label{aaa}
\end{equation}%
The Fourier-Laplace transform of the total density $\rho (x,t)=\rho
_{+}(x,t)+\rho _{-}(x,t)$ is $\tilde{\rho}_{+}(k,s)+\tilde{\rho}_{-}(k,s).$
Using (\ref{aa}) and (\ref{aaa}), we obtain (\ref{full}).

\section{Appendix 2: anomalous switching $\protect\mu <1$.}

In this Appendix we consider the case when the death rate $\theta =0$ and
all cells start at $t=0$ to move on the right with the velocity $v$ from the
point $x=0:$
\begin{equation*}
\rho _{+}^{0}(x)=\delta \left( x\right) ,\quad \rho _{-}^{0}(x)=0.
\end{equation*}%
Then $\rho _{+}^{0}(k)=1$ and $\rho _{-}^{0}(k)=0.$ It follows from (\ref%
{full}) that
\begin{equation*}
\tilde{\rho}(k,s)=\frac{\hat{\Psi}(s-ikv)+\hat{\Psi}(s+ikv)\hat{\psi}(s-ikv)%
}{1-\hat{\psi}(s+ikv)\hat{\psi}(s-ikv)}.
\end{equation*}%
By using this formula, we can find the Laplace transforms of the first
moment $\mathbb{E}\left\{ x(t)\right\} $ as
\begin{equation*}
\mathbb{E}\left\{ x(s)\right\} =\frac{\partial \tilde{\rho}(k,s)}{\partial
\left( ik\right) }|_{k=0}.
\end{equation*}%
When $\mu <1$ the first moment $<T>=\int_{0}^{\infty }\tau \psi (\tau )\tau $
is divergent. We obtain%
\begin{equation*}
\mathbb{E}\left\{ \hat{x}(s)\right\} \simeq \frac{v\Gamma (1-\mu )\tau
_{0}^{\mu }}{2s^{2-\mu }}.
\end{equation*}%
Inverse Laplace transform gives
\begin{equation}
\mathbb{E}\left\{ x(t)\right\} \simeq \frac{v\tau _{0}^{\mu }}{2}t^{1-\mu }.
\label{an}
\end{equation}%
The same anomalous behaviour of the first moment $\mathbb{E}\left\{
x(t)\right\} $ was obtained for the fractional Kramers equation \cite{Barkai}
(see also Appendix 4).

\section{\protect\bigskip Appendix 3: anomalous switching $1<\protect\mu <2$.%
}

In this Appendix we discuss the case when the death rate $\theta =0$ and $%
1<\mu <2.$ The purpose is to show that cell motility exhibits subballistic
superdiffusive behaviour. We consider now the symmetrical initial conditions
for which $\mathbb{E}\left\{ x(t)\right\} =0$. At $t=0$ the cells start to
move from the point $x=0$ as follows
\begin{equation*}
\rho _{+}^{0}(x)=\frac{1}{2}\delta \left( x\right) ,\quad \rho _{-}^{0}(x)=%
\frac{1}{2}\delta \left( x\right) .
\end{equation*}%
Their Fourier transforms are equal: $\rho _{+}^{0}(k)=\rho _{-}^{0}(k)=1/2.$
Let us find the mean square displacement $\mathbb{E}\left\{ x^{2}(t)\right\}
$. The formula for $\tilde{\rho}(k,s)$ is
\begin{equation}
\tilde{\rho}(k,s)=\frac{\left( 1+\hat{\psi}(s+ikv)\right) \hat{\Psi}%
(s-ikv)+\left( 1+\hat{\psi}(s-ikv)\right) \hat{\Psi}(s+ikv)}{2\left[ 1-\hat{%
\psi}(s+ikv)\hat{\psi}(s-ikv)\right] }  \label{ge}
\end{equation}%
which was firstly obtained by CTRW formalism (see Eq. (9) together with (11)
in \cite{Mas}). One can find the Laplace transform of the second moment $%
\mathbb{E}\left\{ x^{2}(t)\right\} $ using $\tilde{\rho}(k,s)$ from (\ref{ge}%
) as
\begin{equation}
\mathbb{E}\left\{ x^{2}(s)\right\} =\frac{\partial ^{2}\tilde{\rho}(k,s)}{%
\partial \left( ik\right) ^{2}}_{|k=0}=\frac{2v^{2}}{s^{3}}+\frac{4v^{2}\hat{%
\psi}^{\prime }(s)}{s^{2}\left( 1-\hat{\psi}^{2}(s)\right) }.  \label{second}
\end{equation}%
We consider the switching rate (\ref{inverse}) with $1<\mu <2$ when the
first moment $<T>=\int_{0}^{\infty }\tau \psi (\tau )\tau $ is finite, while
the variance is divergent. The small $s$ expansion of $\hat{\psi}\left(
s\right) $ can be written as
\begin{equation}
\hat{\psi}\left( s\right) \simeq 1-<T>s+A<T>s^{\mu }.  \label{small}
\end{equation}%
Substitution of (\ref{small}) into (\ref{second}) gives
\begin{equation*}
\mathbb{E}\left\{ x^{2}(s)\right\} \simeq \frac{2A\left( \mu -1\right) v^{2}%
}{s^{4-\mu }}.
\end{equation*}%
This formula allows us to find the mean squared displacement $\mathbb{E}%
\left\{ x^{2}(t)\right\} $ which exhibits subballistic superdiffusive
behaviour \cite{Mas}
\begin{equation*}
\mathbb{E}\left\{ x^{2}(t)\right\} \simeq \frac{2A\left( \mu -1\right) v^{2}%
}{\Gamma (4-\mu )}t^{3-\mu }.
\end{equation*}%
as $t\rightarrow \infty .$

\section{Appendix 4: velocity autocovariance and mean cell position}

The purpose of this Appendix is to find the mean cell position and to show
that the cell velocity has a long memory for $\mu <1$. Let the cell's
velocity at the initial time be positive, $v(0)=v,$ then the velocity $v(t)$
and the position $x(t)$ of cell can be defined as%
\begin{equation}
v(t)=(-1)^{N(t)}v,  \label{vel}
\end{equation}%
\begin{equation}
x(t)=v\int_{0}^{t}(-1)^{N(u)}du,  \label{pos}
\end{equation}%
where $N(t)$ is the random number of switching up to time $t$ \cite{Kac}.
Autocovariance $C_{v}(t)=\mathbb{E}\left( v(t)v(0)\right) $ and the mean
cell position $\mathbb{E}\left( x(t)\right) $ can be found as

\begin{equation}
C_{v}(t)=v^{2}\mathbb{E}\left\{ (-1)^{N(t)}\right\} =v^{2}\sum_{n=0}^{\infty
}(-1)^{n}P(n,t),
\end{equation}%
\begin{equation}
\mathbb{E}\left\{ x(t)\right\} =v\mathbb{E}\left\{
\int_{0}^{t}(-1)^{N(u)}du\right\} =v\sum_{n=0}^{\infty
}(-1)^{n}\int_{0}^{t}P(n,u)du,  \label{me}
\end{equation}%
where $P(n,t)=\Pr (N(t)=n).$ We should note that in the anomalous case $\mu
<1,$ the random velocity $v(t)$ is non-stationary process. The Laplace
transform of $C_{v}(t)$ and $\mathbb{E}\left( x(t)\right) $ are
\begin{equation}
\hat{C}_{v}(s)=v^{2}\sum_{n=0}^{\infty }(-1)^{n}\hat{P}(n,s),  \label{La0}
\end{equation}%
\begin{equation}
\mathbb{E}\left\{ \hat{x}(s)\right\} =v\sum_{n=0}^{\infty }(-1)^{n}\frac{%
\hat{P}(n,s)}{s},  \label{LL1}
\end{equation}%
where the Laplace transform of $P(n,t)$\ is given by \cite{Feller}
\begin{equation}
\hat{P}(n,s)=\frac{\tilde{\psi}^{n}(s)(1-\tilde{\psi}(s))}{s}.  \label{La1}
\end{equation}%
The substitution of (\ref{La1}) into (\ref{La0}) gives
\begin{equation*}
\hat{C}_{v}(s)=\frac{v^{2}(1-\tilde{\psi}(s)}{s}\sum_{n=0}^{\infty }(-1)^{n}%
\tilde{\psi}^{n}(s)=\frac{v^{2}(1-\tilde{\psi}(s)}{s\left( 1+\tilde{\psi}%
(s)\right) }.
\end{equation*}%
When the mean waiting time $\langle T\rangle =\int_{0}^{\infty }\tau \psi
(\tau )d\tau $ is infinite, the Laplace transform $\tilde{\psi}(s)$ can be
approximated for small $s$ by Eq. (\ref{anom}). In this case we obtain

\begin{equation*}
\hat{C}_{v}(s)\simeq \frac{v^{2}\Gamma (1-\mu )\tau _{0}^{\mu }}{2s^{1-\mu }}%
,
\end{equation*}%
\begin{equation*}
\mathbb{E}\left\{ \hat{x}(s)\right\} \simeq \frac{v\Gamma (1-\mu )\tau
_{0}^{\mu }}{2s^{2-\mu }}.
\end{equation*}%
Inverse Laplace transform gives the large time asymptotics for $0<\mu <1:$
\begin{equation*}
C_{v}(t)\simeq \frac{v^{2}\tau _{0}^{\mu }}{2t^{\mu }},\;\;
\end{equation*}%
\begin{equation*}
\mathbb{E}\left\{ x(t)\right\} \simeq \frac{v\tau _{0}^{\mu }}{2}t^{1-\mu }.
\end{equation*}

\end{document}